\documentclass[
reprint,
superscriptaddress,
onecolumn,
 amsmath,amssymb,
 aps,
floatfix,
]{revtex4-2}

\usepackage{graphicx}% Include figure files
\usepackage{dcolumn}% Align table columns on decimal point
\usepackage{bm}% bold math
\usepackage{xcolor}
\usepackage{hyperref}% add hypertext capabilities
\hypersetup{colorlinks=true,citecolor=blue,linkcolor=red,urlcolor=blue}
\usepackage{bigints} %big integral sign
\usepackage{placeins}
\newcommand{\unimi}[0]{Department of Physics, Center for Complexity and Biosystems, University of Milan, Via Celoria 16, 20133 Milan, Italy}

\begin{document}

% \preprint{APS/123-QED}

\title{Fracture toughness and auxeticity in disordered metamaterials}

\author{Hannes Holey}
\affiliation{\unimi}

\author{Andrea Lorenzo Henri Sergio Detry}
\affiliation{\unimi}

\author{Silvia Bonfanti}
\affiliation{\unimi}
\affiliation{NOMATEN Centre of Excellence, National Center for Nuclear Research, Swierk/Otwock, Poland}

\author{Roberto Guerra}
\affiliation{\unimi}

\author{Anshul D. S. Parmar}
\affiliation{NOMATEN Centre of Excellence, National Center for Nuclear Research, Swierk/Otwock, Poland}

\author{Jacopo Fiocchi}
\affiliation{CNR - Consiglio Nazionale delle Ricerche, Istituto di Chimica della Materia Condensata e di Tecnologie per l'Energia, Via G.Previati 1/E, 23900 Lecco, Italy}

\author{Ausonio Tuissi}
\affiliation{CNR - Consiglio Nazionale delle Ricerche, Istituto di Chimica della Materia Condensata e di Tecnologie per l'Energia, Via G.Previati 1/E, 23900 Lecco, Italy}

% + others

\author{Michael Zaiser}
\affiliation{Institute of Materials Simulation, Department of Materials Science Science and Engineering, Friedrich-Alexander-University Erlangen-Nuremberg, Dr.-Mack-Str. 77, 90762, F{\"u}rth, Germany}

\author{Stefano Zapperi}%
\affiliation{\unimi}
\affiliation{CNR - Consiglio Nazionale delle Ricerche, Istituto di Chimica della Materia Condensata e di Tecnologie per l'Energia, Via R Cozzi 53, 20125 Milan, Italy}
\email{stefano.zapperi@unimi.it}

\date{\today}% It is always \today, today,

\begin{abstract}
Auxetic metamaterials are commonly thought to exhibit favorable mechanical properties, notably high energy absorption. Here we investigate disordered metamaterials obtained from random beam networks by optimizing simultaneously auxeticity and the energy absorbed before fracture. By giving different weights to these optimization targets, we demonstrate that the optimal configurations are connected along a Pareto front where high auxeticity implies comparatively low energy absorption and vice versa. We study the mechanical properties of the resulting metamaterials and characterize the different deformation modes obtained for distinct optimization targets. The simulation and optimization results are validated by comparison with the deformation behavior of additively manufactured samples. Our work provides an illustration of the potentials and limitations of multi-objective optimization in the design of disordered mechanical metamaterials.  
\end{abstract}

\maketitle

%\tableofcontents

\section{Introduction}
Auxetic materials and structures display the uncommon  property of expanding orthogonally when stretched 
resulting in a negative Poisson's ratio. Beside their unusual mechanical response, these materials 
display other potentially useful mechanical properties, such as enhanced fracture and indentation resistance 
which might be generically associated with auxeticity (see e.g.\ \cite{mir2014review}).
On the continuum level, this association appears to be reasonable since strong lateral contraction in the tensile zone ahead of a crack tip is likely to promote crack closure, preventing cracks from advancing \cite{carneiro2013auxetic}. Because of this, auxetic materials are often considered for a variety of applications ranging from shock and impact protection devices \cite{evans1991auxetic,evans2000_auxetic} 
to bioprostheses \cite{scarpa2008auxetic} and aerospace components \cite{alderson2007auxetic}.  These applications often require fine tuned mechanical properties which could be achieved by a careful design of the structure in the framework of flexible mechanical metamaterials \cite{bertoldi2017flexible}.

Auxetic metamaterials are often constructed by periodic repetition of a simple unit cell, resulting in highly ordered structures such as the bowtie lattice, where the negative Poisson ratio is due to the re-entrant cellular structure \cite{evans1995auxetic}. Improved auxetic behavior may be achieved by applying mathematical methods such as topology optimization to these unit cells \cite{schwerdtfeger2011design}. At the same time, it must be emphasized that auxetic behavior on the macroscale does not guarantee favorable failure properties, as the actual failure process may be controlled by correlated modes of cell collapse or cell wall fracture that reflect processes on the intra-cellular rather than the macroscopic level. Recently, it has been argued that the consideration of disordered rather than ordered metamaterials might offer generic pathways to avoid such correlated modes of local failure \cite{zaiser2023_disordered}. 

Two general approaches for the design of disordered auxetic metamaterials have been discussed in the literature. Hanifpour \textit{et al.} \cite{hanifpour2018_mechanics} started from ordered auxetic structures consisting of periodically arranged cells and then introduced increasing random perturbations into the cell arrangement. This approach preserves information about the underlying periodic structure. As a consequence, while the negative Poisson's number values are remarkably robust under perturbation, the same is true for the correlated failure modes for which the local perturbations offer favorable initiation sites. Thus, a moderate level of random perturbations leads to a decrease of strength at approximately constant auxetic properties, without major changes in failure mode. Large perturbations, on the other hand, deteriorate both strength and auxeticity.

An alternative approach for the construction of random auxetic structures relies on computational-assisted design \cite{bonfanti2024computational}. An example is the work of Reid \textit{et al.} on random spring networks \cite{reid2018_auxetic}, later expanded by Shen \textit{et al.} \cite{shen2024autonomous} to include bending and torsional degrees of freedom. This method is based on the construction of an initial, non auxetic random network of springs whose mechanical properties are then modified by selective removal of springs. Such approach has been used to construct auxetic metamaterials \cite{reid2018auxetic} and to tune random spring networks to obtain metamaterials with improved impact properties \cite{reyes-martinez2022_tuning}. The target in these studies was focused only on the auxeticity of the structure without explicitly considering fracture or energy absorption, which would then result only as a by-product of the optimization. Similar optimization strategies, based on simulated annealing of elastic networks endowed with bending and torsional degrees of freedoms, were exploited to to design 
mechanical metamaterial actuators \cite{bonfanti2020_automatic} and chiral metamaterials \cite{beretta2021automatic}

In the present investigation, we conduct a systematic study of random metamaterials obtained by stochastic optimization of random beam networks  through a simulated annealing approach. Beam networks  have an advantage over previously studied spring networks \cite{bonfanti2020_automatic,beretta2021automatic,shen2024_autonomous}, because elastic deformation can readily be obtained through a linear solver and 
the fracture process can easily be simulated \cite{herrmann1989fracture,alava2006statistical,nukala2008crack}.
It is our aim to establish the relationship between auxetic behavior and fracture toughness in this class of materials, by conducting multi-objective optimization in the domain of Poisson's ratio and fracture energy, in view of exploring favorable combinations. Construction and optimization of random network structures (Section \ref{sec:methods}) is accompanied by analysis of their mechanical properties. These properties are studied on the macroscopic level (fracture energy and Poisson's ratio) as well as on the level of local and global stress distributions, deformation and failure modes, and in terms of network theoretical measures such as edge centrality. The simulation studies are accompanied by experimental fabrication of samples via additive manufacturing and experimental validation of the computed properties (Section \ref{sec:results}).

\section{Methods}\label{sec:methods}

\subsection{Network generation}\label{methods:networks}

Initial 2D beam networks were derived from a dense and highly stable random packing of polydisperse disks constructed following the method of Berthier~\textit{et al.} ~\cite{berthier2016equilibrium,parmar2020ultrastable}. Once such a packing is constructed, we identified the nodes of the network (the beam connection points) with the disk centers and the edges with beams that connect the centers of disks that are in mutual contact. The rationale behind this procedure is to obtain a random network that is both dense and homogeneous (without the long-range density fluctuations, e.g.\ associated with a Voronoi network), and at the same time has no underlying periodic structure assumed in other methods, see e.g.\ \cite{hanifpour2018_mechanics,liebenstein2018size}.

To obtain the underlying disk packing, we simulated a discrete-polydisperse mixture of \( N = 2000 \) soft disks distributed across \( M = 20 \) species, with diameters $\sigma$ drawn from the distribution \( P(\sigma) \propto \sigma^{-3} \), over the range \( \sigma \in [0.73, 1.62] \). The mean diameter \( \langle \sigma \rangle \) was taken as the unit of length. We assumed two disks $i$ and $j$ to interact if their distance $r_{ij}$ is less than their mean diameter $\sigma_{ij} = (\sigma_i + \sigma_j)/2$. The pairwise interaction potential is given by \cite{chaudhuri2010jamming}
\begin{equation}
    e(r_{ij}) = \frac{\epsilon}{2}\left(1- \frac{r_{ij}}{\sigma_{ij}}\right)^2 + \sum_{k=0}^2 c_{2k}\left(\frac{r_{ij}}{\sigma_{ij}}\right)^{2k},\;r_{ij}\leq\sigma_{ij},
\end{equation}
with $c_0=-\epsilon/8$, $c_2=\epsilon/4$, and $c_4=-\epsilon/8$. 
We considered a periodic simulation cell of size $L \times L$ and defined the packing fraction as
\[
\phi = \sum_{i=1}^N \frac{\pi \sigma_i^2}{4L^2}.
\]
In a first step, liquid configurations were equilibrated via swap Monte Carlo at \( \phi = 0.78 \)~\cite{berthier2016equilibrium}. The system was then compressed in steps of \( \Delta L = 10^{-3} \) until the non-dimensional pressure reached \( 10^{-6} \), and subsequently decompressed until the pressure dropped below \( 10^{-16} \), yielding jammed states at \( \phi \approx 0.853 \)~\cite{chaudhuri2010jamming}, above the random close-packing threshold (\( \phi \approx 0.845 \)). At each compression and decompression step, energy minimization was performed using the conjugate gradient method, ensuring the jammed packing to be in mechanical equilibrium. Finally, the jammed packing was scaled to the desired system size. 

The disordered-jammed packing of spheres was then converted into a beam network by connecting the centers of contacting particle pairs. Two particles were defined to be in contact if $(r_{ij} - \sigma_{ij}) \in [0,  10^{-3} \langle \sigma \rangle]$. Beams that crossed the boundaries of the periodic simulation cell were removed to obtain a free standing network of approximately square shape. 

\subsection{Timoshenko beam model}\label{methods:beams}

We modeled the samples as planar networks of $N_e$ rigidly connected Timoshenko beams. The network was assumed to be contained in the $x$-$y$ plane with unit normal $\mathbf{n} \equiv \mathbf{e}_z$.
%\RG{the $x$-$y$ plane with unit normal $\mathbf{n} \equiv z$}. 
%
Thus, in two dimensions, each node $i$ was associated with two translational degrees of freedom represented by a displacement vector $\mathbf{u}_i$ with $\mathbf{u}_i \cdot \mathbf{n} = 0$, and one rotational angle $\theta_i$ describing the rotation of the node around $\mathbf{n}$. 
We considered a single beam element connecting nodes $i$ located at $\mathbf{r}_i$ and $j$ located at $\mathbf{r}_j$. The connecting vector is denoted as $\mathbf{s}_e = \mathbf{r}_j - \mathbf{r}_i$ and its length is denoted as $l_{e} = |\mathbf{s}_e|$. A local element frame is spanned by the unit vectors $\mathbf{l}_{e} = \mathbf{s}_e/l_e$ and $\mathbf{m}_{e} = \mathbf{n} \times\mathbf{l}_{e}$.

The constitutive behavior of the element is specified in the local element frame. The degrees of freedom of the beam element are collected into the generalized displacement vector $\tilde{\mathbf{u}}=\left(\tilde{u}_i, \tilde{v}_i, \theta_i, \tilde{u}_j, \tilde{v}_j, \theta_j\right)^\top$, where $\tilde{u}_i$ and $\tilde{v}_i$ are the components of $\mathbf{u}_i$ in the local element frame.
The local element stiffness matrix for a Timoshenko beam element is in the same frame given by
\begin{equation}
{\renewcommand{\arraystretch}{2.0} %increases the vertical spacing in the matrix
\tilde{\mathbf{K}}_e=\begin{pmatrix}
\frac{EA}{l_e}& 0 & 0 & -\frac{EA}{l_e} & 0 & 0 \\
& \frac{12E I_z}{l_e^3(1+\Phi_y)} & \frac{6E I_z}{l_e^2(1+\Phi_y)} & 0 & -\frac{12E I_z}{l_e^3(1+\Phi_y)} & \frac{6E I_z}{l_e^2(1+\Phi_y)} \\
& & \frac{(4+\Phi_y) E I_z}{l_e(1+\Phi_y)} & 0 & - \frac{6E I_z}{l_e^2(1+\Phi_y)} & \frac{(2-\Phi_y) E I_z}{l_e(1+\Phi_y)} \\
& & & \frac{EA}{l_e} & 0 & 0 \\
& & & & \frac{12E I_z}{l_e^3(1+\Phi_y)} & -\frac{6E I_z}{l_e^2(1+\Phi_y)} \\
\text{sym.} & & & & & \frac{(4+\Phi_y) E I_z}{l_e(1+\Phi_y)}
\end{pmatrix},
}
\label{eq:local_stiffness}
\end{equation}
with
\begin{equation}
\Phi_y=\frac{12 E I_z}{\kappa G A l_e^2}.
\label{eq:local_stiffness_phi}
\end{equation}
In Eqs.~\eqref{eq:local_stiffness} and ~\eqref{eq:local_stiffness_phi}, $E$ denotes the beam's Young's modulus, $G$ its shear modulus, and $\kappa$ is the shear correction factor for different cross section shapes. $A$ and $I_z$ are the beam cross section and its second moment of area, respectively, which we assumed to be identical for all beams. In the following, we consider beams with rectangular cross section, where the width of the beam in $\mathbf{m}$ direction is denoted as $d_m$ and its width in $\mathbf{n}$ direction as $d_n$, hence $A = d_md_n$ and $I_z = d_m^3 d_n/12$.
To model the collective response of the beam network, we transformed the element stiffness matrix into a global frame with unit vectors $\mathbf{e}_x$ and $\mathbf{e}_y$, i.e. $\mathbf{K}_e=\mathbf{T}_e^\top\tilde{\mathbf{K}}_e\mathbf{T}_e$, where $\mathbf{T}_e$ are element transformation matrices given by the direction cosines of the beam coordinate system, i.e.,
\begin{equation}
{\renewcommand{\arraystretch}{1.5} %increases the vertical spacing in the matrix
\mathbf{T}_e=\begin{pmatrix}
\mathbf{l}_{e} \cdot \mathbf{e}_x & 
\mathbf{l}_{e} \cdot \mathbf{e}_y & 0  \\
\mathbf{m}_{e} \cdot \mathbf{e}_x & 
\mathbf{m}_{e} \cdot \mathbf{e}_y & 0  \\
0 & 0 & 1  \\
\mathbf{l}_{e} \cdot \mathbf{e}_x & 
\mathbf{l}_{e} \cdot \mathbf{e}_y & 0  \\
\mathbf{m}_{e} \cdot \mathbf{e}_x & 
\mathbf{m}_{e} \cdot \mathbf{e}_y & 0  \\
0 & 0 & 1  \\
\end{pmatrix},
}
\end{equation}
To solve for the generalized displacements  $\mathbf{u}$ in the global frame, we assembled the global stiffness matrix $\mathbf{K}$ by summation over the elements, $\mathbf{K}=\sum_{e=1}^{N_e} \mathbf{L}_e^\top \mathbf{K}_e \mathbf{L}_e$ with element localization matrices $\mathbf{L}_e$.
The network connectivity determines the element localization matrices $\mathbf{L}_e$ which are represented by a block matrix of size $6\times 6N_e$ whose only nonzero blocks are 3$\times$3 unit matrices at $L_{1i}$ and $L_{2j}$, where $i$ and $j$ are the node indices connected by element $e$.
We considered deformation and failure to occur under quasi-static conditions. Hence, balance of linear and angular momentum are described by the linear system of equations
\begin{equation}
\mathbf{K}\cdot\mathbf{u} = \mathbf{f}
\end{equation}
which we solved in the global frame. The external force vector $\mathbf{f}$ and boundary conditions are specified in the next section. 

\subsection{Fracture simulations}\label{methods:fracture}

To complete our constitutive framework for elastic-brittle behavior, 
we considered a fracture criterion where irreversible failure occurs once the von Mises equivalent stress at the most highly loaded cross section of a beam exceeds a critical value, as it is customary in the study of random beam models \cite{alava2006statistical},  In our two-dimensional setting, the von Mises equivalent stress is given by 
\begin{equation}
    \sigma_\mathrm{eq} = \sqrt{\sigma_{11}^2 +  3\sigma_{12}^2}.
\end{equation}
General deformation of the beam involves shear, axial stretch, and bending. The tensile stress is given by $\sigma_{11}(\tilde{x},\tilde{y})=F/A + M(\tilde{x}) \tilde{y} / I$ and the shear stress is $\sigma_{12}=Q/A$, where $F$ and $Q$ are axial and shear forces, respectively, and $M(\tilde{x})$ is the bending moment. $\tilde{x}$ is the local coordinate in $\mathbf{l}$ direction and $\tilde{y}$ is the coordinate in $\mathbf{m}$ direction.
Forces and moments were measured at the beam end nodes $i$ and $j$. In quasi-static equilibrium, the nodal forces form couples. In a small strain approximation (note that strains are assumed small on the beam level but not globally), we fibd that $F_i = - F_j$ and $Q_i = -Q_j$. 
The bending moment $M(\tilde{x})$ varies linearly between the two end nodes. This implies that the critical location for beam failure is at the beam endpoint with the larger absolute value of the bending moment, on the outermost fiber of the beam on the side where the bending stress is tensile. Thus, 
for the failure criterion $\sigma_\mathrm{eq}=t$, the maximum equivalent stress along the beam is given by
\begin{equation}
    \sigma_\mathrm{eq} = \sqrt{\left(\frac{F_i n_i}{A} + \max(|M_i|, |M_j|)\frac{d_m}{2 I_z}\right)^2 + 3\left(\frac{Q_i}{A}\right)^2}.
\end{equation}

We used a quasi-static displacement-controlled fracture protocol.
The basis vectors $\mathbf{e}_x$ and $\mathbf{e}_y$ of the global coordinate system were aligned with the edges of the specimen, whole the sides of the specimen at $x = -L/2$ and $x = L/2$ were free. Nodes within a distance $\delta y = 0.05L$ from the bottom plane $y = -L/2$ were fixed in $y$ direction, and the node closest to the point $(0,0)$ was fixed in both $x$ and $y$ directions to ensure well-posedness of the system of equations. Loading was imposed by displacing nodes within $\delta y$ from the top plane $y = L/2$ rigidly in the upward direction. The upward displacement was increased in discrete steps $v_n^\mathrm{app}$, starting with an initial ``virtual'' step $v^\mathrm{app}_0 = 1$.

Subsequent actual displacements were evaluated using the linearity of the problem, by scaling the applied displacement $v_{k+1}^\mathrm{app}= v_k^\mathrm{app} t/\max(\sigma_{\mathrm{eq}, k})$ in each increment based on the previous increment, such that the maximally stressed beam reaches the failure threshold $t$ \cite{alava2006statistical}. In other words, we set the displacement to the exact level where the beam with the lowest strength margin fails. 
After the removal of this beam, we accounted for stress redistribution due to the changed network topology, and subsequently removed all beams with $\sigma_\mathrm{eq} \geq t$ before proceeding with the next displacement increment.
We measured the reactive forces at the constrained nodes to compute load displacement curves for each fracture simulation. The simulation was terminated upon global failure, as indicated by loss of well-posedness of the momentum balance equations.

To achieve printable designs, we set the minimum allowable beam width to the nozzle diameter of the 3D printer (see Sec.~\ref{methods:exp}), i.e. $d_m\geq0.5\,\mathrm{mm}$, and $d_n=5\,\mathrm{mm}$, where the build direction is aligned with the normal vector $\mathbf{n}$.
Further, we ensured a minimum beam aspect ratio $l_e/d_m\geq5$ by rescaling the nodal coordinates, which leads to a lateral size of $L\approx 150\,\mathrm{mm}$. 
All optimization runs were conducted for beams with Young's modulus $E=3\,\mathrm{GPa}$, shear modulus $G=1.15\,\mathrm{GPa}$, and failure strength $t=0.02E$. 
These parameters were later adjusted for comparison with experimental force displacement curves.

\subsection{Monte Carlo optimization}\label{methods:opt}

We performed Monte Carlo optimization to maximize both the negative Poisson's ratio and the fracture toughness at constant mass.
In the trial step, we removed or added bonds at random locations from the initial network structure while keeping the mass of the system fixed, i.e., we either add  $\pm d_{m,0}/N_e$ to all beam widths in $\mathbf{m}$ direction, or $\pm d_{n,0}/N_e$ to all beam widths in $\mathbf{n}$ direction. These two variants are not equivalent. Rescaling the width in $\mathbf{n}$ direction increases the axial and bending stiffness in equal proportions, since both $A$ and $I$ are linear in $d_n$. Rescaling the beam width in $\mathbf{m}$ direction increases the bending resistance relative to the axial stiffness because of the cubic scaling of the beam's second moment of area with $d_m$.

After each trial step, we performed a quasi-static fracture simulation and computed the Poisson's ratio $\nu=-d\varepsilon_x / d\varepsilon_y$ from a small initial load step.  For the calculation of the transverse strain increment $d\varepsilon_x$, we divided the sample into bins in $x$-direction, and averaged the nodal $x$-displacements within a bin to compute the discrete displacement gradient in $x$ direction, which we averaged over all bins.
Furthermore, we evaluated the energy absorbed during the fracture simulation, $U_\mathrm{frac}=\int_0^{v^\mathrm{app}_\mathrm{fail}} F_y^\mathrm{top} dv^\mathrm{app}$, where $F_y^\mathrm{top}$ is the sum of the reactive forces in $y$ direction at the constrained nodes within $\delta y$ from the top plane $y=L/2$  (see Sec.~\ref{methods:fracture}).

The two optimization objectives were finally combined into a single multi-objective cost function.
A multi-objective optimization problem can generally be written as \citep{giagkiozis2015_methods}
\begin{equation}
\min_\mathbf{x} g(f_1 (\mathbf{x}), \ldots, f_M (\mathbf{x})),
\end{equation}
where $f_i(\mathbf{x}),\;i=1\ldots M$ are the individual objectives and $\mathbf{x}$ is the decision vector.
Here, the decision vector takes binary values and has the length $N_\mathrm{e}$, where $N_\mathrm{e}$ is the number of initial beam elements.
A common approach is to apply \emph{weighted metrics methods}, which use a scalarization function of the form
\begin{equation}
\min_\mathbf{x} \left(\sum_{i=1}^M w_i f_i(\mathbf{x})^p\right)^\frac{1}{p},
\end{equation}
where $w_i$ denote weighting parameters with $w_i \geq 0$ and $\sum_i w_i = 1$.
Here, we used the limiting case of $p=\infty$, which leads to the so-called Chebyshev scalarizing function
\begin{equation}
\min_\mathbf{x} \left(\max\{w_1 f_1(\mathbf{x}), \ldots, w_N f_M(\mathbf{x})\}\right).
\end{equation}
The $M=2$ single-objective functions are $f_1 = \nu/(1+\nu_\mathrm{ref})-(1+2\nu_\mathrm{ref})/(1+\nu_\mathrm{ref})$ and $f_2 = -U_\mathrm{frac}/U_{0}$, where $\nu_\mathrm{ref}$ is a reference Poisson's ratio, and $U_{0}$ is the fracture energy of the initial random structure.
We set $\nu_\mathrm{ref}=0.6$ for all samples, which corresponds approximately to the Poisson's ratio of the initial random structures.
The above scaling ensures that the individual objectives for fracture energy and Poisson's ratio have similar magnitudes.
Weights were selected from the range $w_1\in[0.3, 0.7]$.
The optimization followed the standard Metropolis algorithm, where a trial step ($n + 1$) was accepted for $\Delta=g_{n+1} - g_n < 0$, otherwise it was accepted with probability $P=\exp\left(-\Delta/T\right)$.
The parameter $T$ acted as temperature and we used a simulated annealing protocol \cite{bonfanti2020_automatic,beretta2021automatic}, in which the temperature was reduced exponentially from $T=0.072$ to $10^{-9}$ within the first 100 accepted steps.
After the initial annealing phase, only improvements of the objective functions were accepted.

\subsection{Experimental program}\label{methods:exp}

% All samples shared identical external dimensions but differed in their internal geometry as a result of the computational optimization procedure described in Section~\ref{methods:opt}.
%
We performed mechanical tests of additively manufactured specimen for a single representative structure selected from the samples with rescaled $\mathbf{n}$ direction (due to the higher printing resolution in that direction).
In total, we considered four replicas of the optimized configuration (OPT\_1 to OPT\_4) and two of the non-optimized configuration (NON\_OPT\_1 and NON\_OPT\_2).
%
% Four optimized structures (OPT\_1 to OPT\_4) were selected from the Pareto front, while two unoptimized ones (Initial 1 and Initial 2) served as reference.
% 
The specimens were fabricated using the Step 3.0 3D printer (Oniro S.r.l., Italy), employing a foam additive manufacturing process. The samples were printed in foamed PLA using the Oniro Step 3.0 system. According to the manufacturer (FiloAlfa), the base material (PLA) has a density of 1.24 g/cm³ and a flexural modulus of 3.8 GPa. The tensile modulus is reported as 3.6 GPa, while the tensile strength is 53 MPa. Final printed samples had external dimensions of approximately 150 mm × 150 mm, excluding the gripping tabs.

To ensure stable and reproducible clamping, each sample was printed with integrated gripping tabs at both ends. These tabs were specifically designed to increase the contact area with the testing machine grips, while keeping the central gauge region free from mechanical interference.
Uniaxial tensile tests were performed using an MTS 2/M universal testing machine (MTS Systems Corporation, USA), equipped with custom flat grips and a 2 KN load cell. All tests were carried out under displacement control at a constant crosshead speed of 0.8 mm/min. Only the gripping tabs were clamped during testing to ensure unconstrained deformation of the sample.
Three key mechanical properties were extracted from the tests: Effective stiffness (calculated from the initial slope of the stress-strain curve), the fracture energy (defined as the area under the force displacement curve), and Poisson’s ratio (obtained from image-based tracking of transverse and axial deformations).

Each test was recorded using a XIMEA xiQ USB3 industrial camera (model MQ013MG-E2), mounted on a fixed tripod orthogonal to the specimen. High-resolution grayscale videos were acquired at 38 frames per second under uniform lighting.  Customized image analysis scripts were used to extract transverse and longitudinal strains from the frame-by-frame evolution of the specimen geometry during a test by digital image correlation. Poisson’s ratio was then computed as the negative ratio between the average transverse and axial strain increments in the elastic regime. Further details on the image-based method are provided in the Supplementary Information
and the code is accessible at \url{https://github.com/andrea199/PossionCalc}.The fracture energy was computed directly from the raw load displacement data acquired during mechanical testing via numerical integration using the trapezoidal rule. 

% After cleaning and rescaling the experimental files, the area under the curve was numerically integrated using the trapezoidal method to obtain the total energy absorbed up to failure.

\section{Results}\label{sec:results}

\subsection{Outcomes of optimization}

\begin{figure}[!t]
\centering
\includegraphics{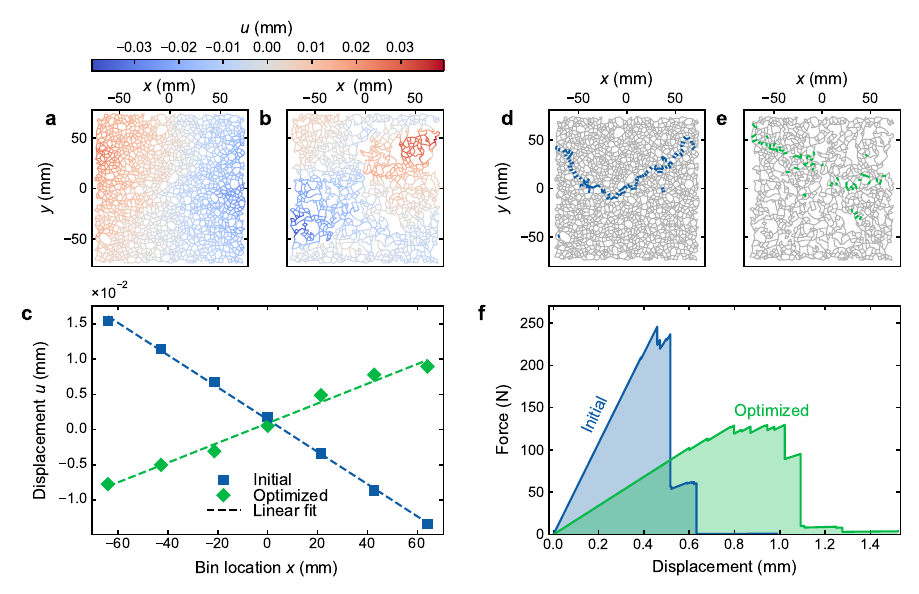}
\caption{%
\textbf{Evaluation of auxeticity and fracture energy of the disordered networks.}
\textbf{a} Initial network structure under uniaxial tension in $y$ direction. The colormap indicates the lateral displacement ($x$) of the edges as averages of the nodal displacements. 
\textbf{b} Optimized structure under uniaxial tension, optimization parameter $w_1=0.61$. 
\textbf{c}  Lateral displacement field in both optimized and final structures. Squares represent bin averages along $x$ and dashed lines are linear fits. Thus, the slope approximates the strain increment $d\varepsilon_x$ (assuming small tensile strain), to calculate the Poisson ratio via $\nu=-d\varepsilon_x/d\varepsilon_y$.
\textbf{d} Failed edges at the end of the fracture process for the initial structure, and, 
\textbf{e} for the optimized structure, highlighted in blue and green, respectively.
% Colored bonds are those broken at the end of the fracturing process.
\textbf{f} Force displacement curves under uniaxial tension for the initial (blue) and optimized (green) structure. The fracture energy is calculated from the shaded area under the curves.}
\label{fig:sim_single}
\end{figure}

In total, we optimized 200 structures for negative Poisson's ratio and high fracture energy with a set of weights sampled uniformly from the interval $w_1\in[0.3, 0.7]$.
Half of the structures were evaluated under tension and the other half under compression.
Figure~\ref{fig:sim_single} illustrates the evaluation of Poisson's ratio and fracture energy for a single sample under tension (in $y$ direction) with optimization parameter $w_1=0.61$, i.e. weighing the auxeticity objective stronger than high fracture energy.

The element-wise beam displacement in $x$ direction is shown in Fig.~\ref{fig:sim_single}a and b for the initial and optimized network, respectively.
While the initial structure shrinks laterally (Fig.~\ref{fig:sim_single}a) as highlighted by the colormap, the optimized structure shows the opposite lateral deformation behavior (Fig.~\ref{fig:sim_single}b), expanding laterally under tension.
Given the discrete, inhomogeneous displacement distribution among the network edges, particularly in the optimized structures, we computed the lateral displacement as bin averages along the $x$ axis.
Fig.~\ref{fig:sim_single}c shows the displacement $u_x$ for seven bins for both initial and optimized structures.
The slope of a linear fit to the displacement bin averages determines the lateral strain $\varepsilon_x$, and thus, with the applied strain $\varepsilon_y$, the Poisson's ratio $\nu=-\varepsilon_x/\varepsilon_y$.
For the example shown here, we obtained $\nu=0.69$ and $\nu=-0.91$ for the initial and optimized network, respectively.

The failed edges during the fracture simulations for initial and optimized structures are highlighted in Fig.~\ref{fig:sim_single}d and e, respectively.
In the initial structures, almost all the failed edges lie along the final crack path, indicating that fracture is dominated by the propagation of a single critical crack.
The failed edges in the optimized structures are more dispersed over the full domain, in line with the idea that auxeticity mitigates against the effect of crack-tip stress concentrations, but the final critical crack is visible as well.
The differences in fracture mechanisms manifest themselves in the load displacement curves shown in Fig.~\ref{fig:sim_single}f. 
The initial structure is approximately three times stiffer than the optimized one but it fails suddenly at approximately $0.4\%$ elongation.
Although much more compliant than the initial structure, the fracture energy of the optimized structure (as given by the area under the force displacement curve) is $25\%$ larger, due to an extended plateau with many ``small'' failure events and a failure strain of approximately $0.8\%$.

\begin{figure}[!t]
\centering
\includegraphics{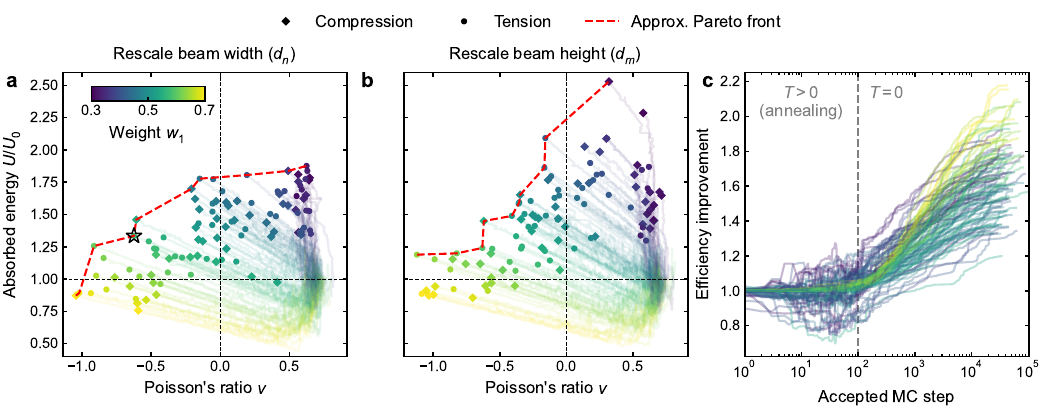}
\caption{%
\textbf{Multi-objective optimization procedure.}
\textbf{a} Results for networks with rescaled beam width during optimization. Diamonds and disks show the final fracture energy and Poisson ratio for samples under compression and tension, respectively. Shaded lines indicate the trajectory during Monte Carlo (MC) optimization. The optimization parameter $w_1$ is represented by the color of the symbols and lines, sampled from the interval $w_1\in[0.3, 0.7]$.
The red dashed line illustrates the approximate Pareto front for all samples and the star highlights the sample selected for experimental validation.
\textbf{b} Same as \textbf{a} but for systems with rescaled beam height.
\textbf{c} Change of efficiency (negative cost function) with respect to its initial value during optimization. The first 100 accepted MC steps were performed at finite temperature following an annealing protocol.
}
\label{fig:sim_all}
\end{figure}

Fracture energy and Poisson's ratio are computed in every Monte Carlo (MC) trial step to evaluate the cost function.  Figure~\ref{fig:sim_all} illustrates the trajectories of accepted Monte Carlo steps for the two different ways of redistributing mass in the network used in this study.
In Fig.~\ref{fig:sim_all} we plot the trajectories in the objective space spanned by the Poisson's ratio and the normalized fracture energy for systems in which the mass difference due to element deletion or addition has been redistributed by scaling of the beam width $d_n$ in the $\mathbf{n}$ direction.
Poisson's ratio for all initial structures is approximately $0.6$.
Samples in which  during optimization the weight assigned to fracture energy was larger than the one given to auxeticity reach up to $80\%$ larger fracture energies than the initial structures with a Poisson's ratio that remains more or less unchanged.
On the other hand, samples optimized with a large weight on the auxeticity objective reach Poisson's ratios $\nu\approx-1$ but show slightly lower fracture energies ($\sim-20\%$) than the initial networks.
The approximate Pareto front (approximate in the sense that we may find better samples if we would sample more) connects the samples for which no improvement in one of the individual objectives can be achieved without deteriorating the other objective.
Although the obtained Pareto front is slightly convex towards the upper left quadrant, our optimized samples do not populate regions of high fracture energy and negative Poisson's ratio.

A similar trend can be observed for structures where mass has been redistributed by scaling the beam thickness $d_m$ in $\mathbf{m}$ direction (Fig.~\ref{fig:sim_all}b).
This method tends to produce large improvements in the fracture energy (up to 150\%) in optimizations where the weight assigned to fracture is stronger than the one assigned to the Poisson's ratio, because the relative increase of the beam bending stiffness during optimization favors stretch-dominated over bending dominated deformation modes. On the other hand, we see little difference between the two mass re-distribution methods when optimization is carried out to obtain high negative Poisson ratio. The overall result is that scaling the beam thickness in $\mathbf{m}$ direction direction produces a more concave Pareto front than scaling along $\mathbf{n}$ direction. Either way, we observe no indication of a synergy between fracture energy and auxeticity. 
Figure~\ref{fig:sim_all}c shows the evolution of the efficiency, i.e. the negative cost, here normalized by the initial efficiency.
During the annealing phase of 100 accepted steps, no substantial improvement can be observed. 
Here, this phase is mostly characterized by deleting beams from the initial structure, thus, exploring the design space before the MC optimization with $T=0$ resumes.
Most samples converged within $10^4$ and $10^5$ accepted steps, leading to efficiency improvements between 20 and 120\%.

\subsection{Global and local mechanical properties}

\begin{figure}[!ht]
\centering
\includegraphics{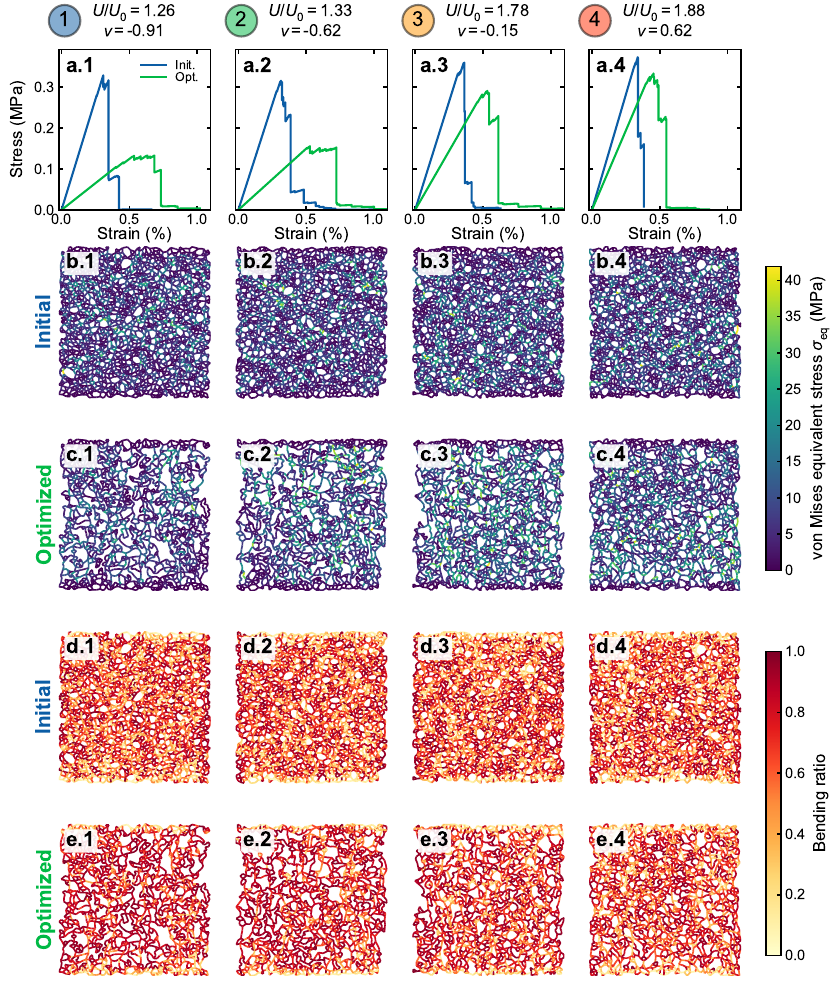}
\caption{%
\textbf{Selected stress strain curves and network structures along the Pareto front.}
\textbf{a} Stress strain curves under uniaxial tension before (``Init.", blue lines) and after optimization (``Opt.", green lines). The four columns 1 -- 4 correspond to samples selected along the Pareto front sorted from low to high fracture energy (or equivalently low to high Poisson's ratio). 
Rows \textbf{b} and \textbf{c} show the von Mises equivalent stress distributions for initial and optimized structures, respectively.
Rows \textbf{d} and \textbf{e} shows the bending ratio $r=\sigma_{11, \mathrm{bend}}/\sigma_{11}$ for the initial and optimized structures, respectively. }
\label{fig:sim_pareto}
\end{figure}

Figure~\ref{fig:sim_pareto} shows stress-strain curves of samples located on the Pareto front, together with the corresponding patterns of local stresses (von Mises stress at beam level) and of the bending ratio $r$, which gives a measure of the degree of bending-dominated versus stretch-dominated deformation at the beam level \cite{rayneau2019density}.
We computed the bending ratio by comparing the bending-induced tensile stress $\sigma_{11, \mathrm{bend}}$ with the total tensile stress $\sigma_{11}$, i.e. $r=\sigma_{11, \mathrm{bend}}/\sigma_{11}$.
In contrast to the failure criterion, we considered the average bending moment at beam level, such that $\sigma_{11, \mathrm{bend}}=d_m(|M_i| + |M_j|)/4 I_z$.
Several trends can be observed. On the level of stress-strain curves, we observe that optimization for high fracture energy leads to a slight increase of the peak stress alongside a very significant increase in failure strain. The overall increase in fracture energy is mainly due to the latter effect. With increasing Poisson ratio, the peak stress decreases but the failure strain remains consistently well above that of the non-optimized samples (Fig.~\ref{fig:sim_pareto}, a.1-4). Figure~\ref{fig:sim_pareto}b.1-4 shows the Von Mises stresses at the beam level computed at 0.2\% strain, and Fig.~ \ref{fig:sim_pareto}c.1-4 shows the same information for the optimized configurations. 

The general trends seen from these figures indicate that optimization for high fracture energy produces dense and homogeneous structures that sustain higher stress levels than structures optimized for auxeticity. In structures optimized for fracture energy, beam deformation is to a larger extent stretch dominated, whereas optimization for auxeticity produces structures where beams to a larger extent deform by bending. 

\begin{figure}[!ht]
\centering
\includegraphics{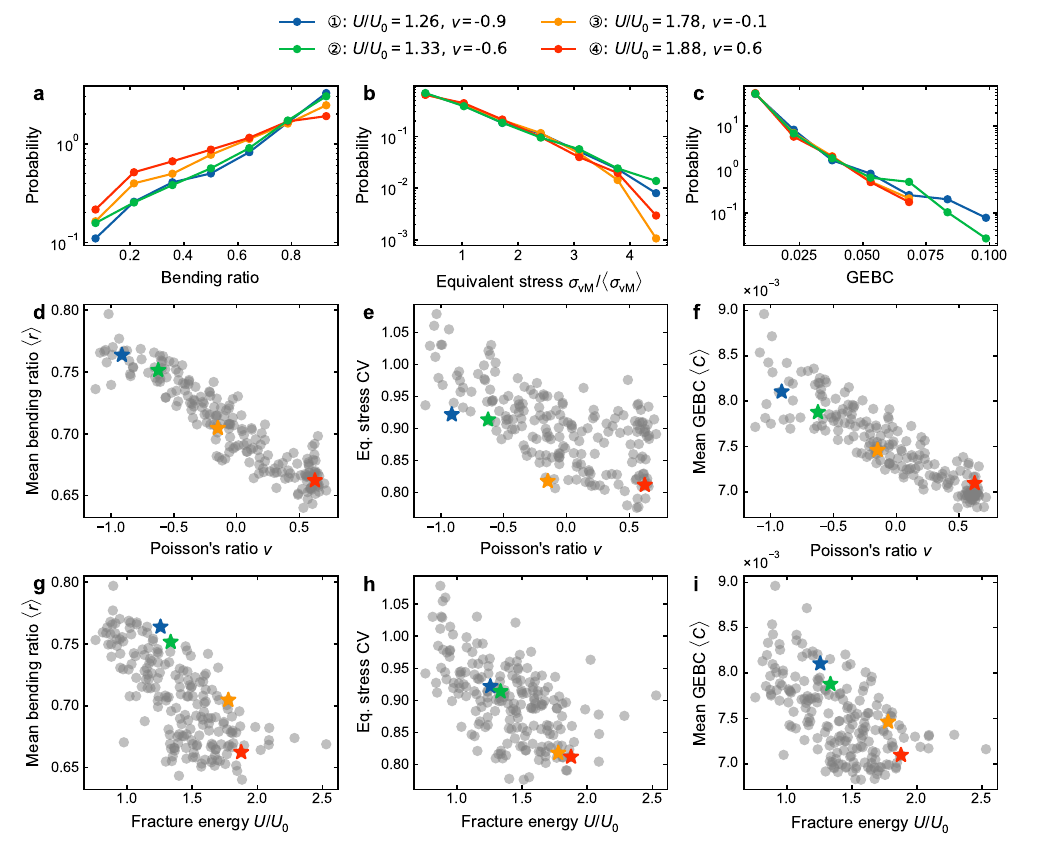}
\caption{%
\textbf{Distribution of beam-level properties for four selected samples along the Pareto front and correlation between moments of these distribution with the optimization targets for all samples.}
\textbf{a} Probability distribution of the bending ratio.
\textbf{b} Probability distribution of the normalized equivalent stress.
\textbf{c} Probability distribution of geodesic edge betweenness centrality (GEBC).
\textbf{d} Mean bending ratio for all samples over Poisson's ratio.
\textbf{e} Coefficient of variation (CV) of the equivalent stress distribution over Poisson's ratio.  
\textbf{f} Mean GEBC over Poisson's ratio.
Panels \textbf{g}--\textbf{i} show the same properties as \textbf{d}--\textbf{f} but plotted over the normalized fracture energy.
Stars in \textbf{d}--\textbf{i} correspond to the selected samples shown in \textbf{a}--\textbf{c} and Fig.~\ref{fig:sim_pareto}.
}
\label{fig:sim_pareto_dist}
\end{figure}

These trends are further borne out by the statistical data shown in Fig.~\ref{fig:sim_pareto_dist}. Figure~\ref{fig:sim_pareto_dist}a and b show, for the four samples illustrated in Fig.~\ref{fig:sim_pareto}, the statistical distributions of Von Mises stresses (normalized by their mean) and of beam-level bending ratios. Figure~\ref{fig:sim_pareto_dist}c shows the distribution of geodesic edge betweenness centrality (GEBC), a network theoretical measure which has been proposed as an indicator of local failure propensity \cite{berthier2019forecasting,moretti2019network,pournajar2022edge}.
The GEBC measures the importance of an edge $e$ within the network by summing the fractions of all shortest paths between two nodes that pass through $e$, i.e.,
\begin{equation}
C(e) = \frac{2}{N(N-1)}\sum_{\substack{i,j\in[1, N]\\ i\neq j}}\frac{n(i, j|e)}{n(i, j)},
\end{equation}
where $n(i, j)$ denotes the number of shortest paths, and $n(i, j | e)$ those that pass through $e$.
Note that GEBC values are normalized by the number of possible unordered node pairs, where $N$ denotes the number of nodes in the network.
We used \textsc{rustworkx} \cite{Treinish2022} to compute the GEBC distribution.
The graphs show that the bending ratio distribution is, in samples optimized for auxeticity, more strongly skewed towards high bending ratios. Also, in these samples the probability of finding high local stresses is higher than in those optimized for high fracture energy, and the same is true for the high-GEBC tail of the edge betweenness centrality distribution. 

The same trends can be found not only along the Pareto front but throughout the entire spectrum of simulated samples. As shown in Fig.~\ref{fig:sim_pareto_dist}d, there is a systematic almost linear correlation between mean bending ratio and Poisson number, such that the strength-optimized specimen with highest Poisson number have the lowest bending ratio and the auxeticity-optimized samples with strongly negative Poisson ratio have the highest bending ratio. 
Similarly, near-linear correlations can be found between the coefficient of variation of the beam stress distribution and the Poisson's ratio as shown in Fig.~\ref{fig:sim_pareto_dist}e, where  strongly negative $\nu$ correlate with wider stress distributions, and between the mean GEBC and the Poisson's ratio as shown in Fig.~\ref{fig:sim_pareto_dist}f, where high mean GEBC correlates with a negative Poisson number.

Systematic but less pronounced correlations can also be observed between the above moments of beam-level properties and the fracture energy. Figure~\ref{fig:sim_pareto_dist}g illustrates that stretch-dominated deformation correlates with high fracture energy, whereas bending-dominated deformation leads to lower fracture energies. Figure~\ref{fig:sim_pareto_dist}h shows that high fracture energy correlates with a low coefficient of variation of the beam stress distribution (in other words, fracture toughness is supported by homogeneous stress states), whereas low fracture energies are associated with high variability of the local stresses. Finally, samples with high mean GEBC values lead to low fracture energies and vice versa, as shown in Fig.~\ref{fig:sim_pareto_dist}i. 

\subsection{Experimental validation}

\begin{figure}[!ht]
\centering
\includegraphics{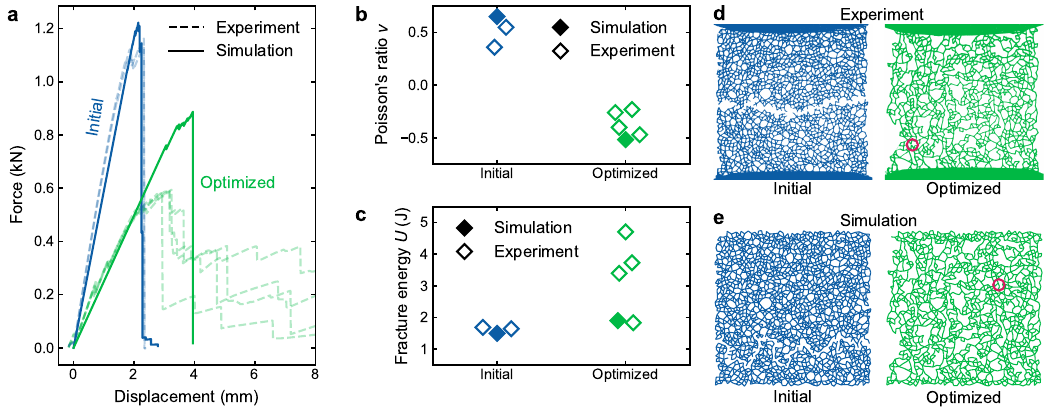}
\caption{%
\textbf{Comparison between experiment and simulation for a single network.}
\textbf{a} Force displacement curve for both initial (blue) and optimized (green) structures for simulations (solid lines) and experiments (dashed lines). Experiments were conducted for two (initial) and four (optimized) replicas of the same network. 
\textbf{b} Poisson's ratio and \textbf{c} fracture energy computed from experimental samples (open symbols) and the simulation (closed symbol).
\textbf{d} Representative crack path and first failure location (red circle) in the initial (left, blue) and optimized (right, green) structures of the experiments, respectively. First failure location was the same in all optimized structures.
\textbf{e} Same as \textbf{d} but for the simulation.
}
\label{fig:sim_vs_exp}
\end{figure}

To validate the modeling approach, experimental tests on additively manufactured samples were used as described in Section~\ref{methods:exp}. To this end, an initial, non-optimized and an optimized structure were considered. 
%\mz{Please indicate the microstructure selected for manufacturing by an asterisk in Figure 2} \hh{Done}.
To assess variability of the manufactured samples that might result from the additive manufacturing process, two samples with the initial and four samples with the optimized microstructure were tested. 
We used one of the non-optimized 3D printed samples to calibrate the Young's modulus and the fracture strength of the simulations to match the slope of the force displacement curve in the elastic regime and the displacement at failure, respectively.
These parameters ($E=3\,\mathrm{GPa}$, $t=250\,\mathrm{MPa}$) were then used to compare simulated force displacement curves also for the optimized structures.
The obtained mechanical characteristics are compiled in Table~\ref{tab:samples_detailed}. A comparison of simulated and experimental samples is shown in Fig.~\ref{fig:sim_vs_exp}. 

\begin{table}[ht]
\centering
\caption{Tested samples and corresponding mechanical properties}
\label{tab:samples_detailed}
\begin{tabular}{lllll}
\hline
\textbf{Sample ID} & \textbf{Type} & \textbf{Poisson's ratio} & \textbf{Young's modulus (GPa)} & \textbf{Fracture energy (J)} \\
\hline
OPT\_1     & Optimized & -0.40 & 0.250 & 4.71 \\
OPT\_2     & Optimized & -0.26 & 0.260 & 3.40 \\
OPT\_3     & Optimized & -0.47 & 0.274 & 1.83 \\
OPT\_4     & Optimized & -0.23 & 0.272 & 3.72 \\
NON\_OPT\_1 & Initial   &  0.36 & 0.669 & 1.68 \\
NON\_OPT\_2 & Initial   &  0.55 & 0.660 & 1.64 \\
\hline
\end{tabular}
\end{table}

As a result of calibration, the simulated force displacement curves for the `initial' samples are in good agreement with the experimental ones, as can be seen in Fig.~\ref{fig:sim_vs_exp}a. For the optimized samples, there is good agreement in the initial elastic regime. Also, the main trend that samples with negative Poisson ratio exhibit lower peak stresses but much larger failure strains, is correctly captured. However, the simulations over-estimated the peak stress while and under-estimated the significant post-peak deformation activity observed in the experimental samples. 

Regarding the optimization parameters, Poisson's ratio and fracture energy, Fig.~\ref{fig:sim_vs_exp}b shows that the experimental values of Poisson's ratio are in good agreement with the theoretical value, while Fig.~\ref{fig:sim_vs_exp}c indicates that the simulated fracture energy represents a lower bound to the experimental data, which exhibit significant statistical variation.

Finally, comparison of Fig.~\ref{fig:sim_vs_exp}d and e indicates that the fracture morphology (crack propagation in the non-optimized samples and diffuse damage accumulation in the optimized ones) is comparable in simulation and experiment. Additional post-fracture images extracted from video frames are shown in the supplementary information. However, neither the fracture pathways nor the points where the first beam failure occurs coincide.  This is likely due to the intrinsic disorder in the microstructure of each beam that is not accounted for in the model where all the beams have the same failure threshold. 

\FloatBarrier

\section{Conclusions}

In this paper, we performed a multi-objective optimization study to investigate the  relationship between auxetic behavior and fracture in disordered metamaterials. Our results indicate that the commonly held opinion that auxetic behavior inherently implies favorable fracture properties, in terms of energy absorbed before failure, does not apply to the disordered metamaterials investigate hered. We find instead that multi-objective optimization, which assigns variable weights to the targets of high fracture energy and low/negative Poisson's ratio, resulted in a Pareto front on which optimized samples show a negative correlation between a high degree of auxeticity and a high fracture energy. 

Investigating the deformation process at the level of individual beams and for large scale internal stress and failure patterns indicates several reasons. The argument that a negative Poisson ratio hinders crack propagation is only of limited relevance to the present structures. Whereas non-optimized samples, which have positive Poisson ratio, fail by crack propagation, the optimized structures show much more diffuse failure patterns associated with pronounced post-peak activity and increased failure strains. 

Our simulation approach produces good overall agreement with experimental results. While agreement in the elastic regime is excellent, there are however some shortfalls in capturing the failure behavior, both in terms of fracture energy, peak stress and failure strain, and in view of the location of failure. This points to the fact that our highly idealized, purely brittle beam failure model may not be fully adequate for capturing the post-peak deformation behavior of the polymeric samples.  

The computational strategy employed here can be easily generalized to other energy absorbing structures requiring to simultaneously control of several mechanical properties at the same time. Energy-absorbing metamaterials 
may have a wide range of advanced applications across several fields due to their ability to efficiently dissipate mechanical energy. In the automotive and aerospace industries, they could provide solutions for crash boxes, bumpers, and cockpit enclosures
thanks to their enhanced toughness. In civil engineering, seismic metamaterials can be embedded in building foundations and joints to mitigate earthquake-induced shocks, protecting infrastructure through localized energy dissipation.

\begin{acknowledgments}
A.D.S.P.\ and S.B.\ are supported by the European Union Horizon 2020 research and innovation program under grant agreement no.\ 857470 and from the European Regional Development Fund via the Foundation for Polish Science International Research Agenda PLUS program grant No. MAB PLUS/2018/8.\ S.B.\ acknowledges support from the National Science Center in Poland through the SONATA BIS grant DEC-2023/50/E/ST3/00569. S.B.\ and A.D.S.P.\ also acknowledge support from the Foundation for Polish Science in Poland through the FIRST TEAM FENG.02.02-IP.05-0177/23 project. J.F., A.T.\ H.H.\ and S.Z.\ were supported by the PRIN 2022 project METACTOR funded from the European Union - Next Generation EU, Mission 4 Component 1 CUP G53D2300164000. H.H.\ also acknowledges the support by the Humboldt foundation through the Feodor-Lynen fellowship. M.Z.\ and S.Z.\ acknowledge support from DFG under Grant No. Za 171 9-3.
R.G.\ acknowledges support from the PRIN 2022 project TRIEL funded by the European Union - Next Generation EU, Mission 4 Component 1 CUP G53D23000790006.
A.L.H.S.D, R.G.\ and S.Z\ acknowledge support from the ARCHIBIOFOAM project which received funding from the European Union’s Horizon Europe research and innovation programme under grant agreement No 101161052. Views and opinions expressed are however those of the author(s) only and do not necessarily reflect those of the European Union or European Innovation Council and SMEs Executive Agency (EISMEA). Neither the European Union nor the granting authority can be held responsible for them.
\end{acknowledgments}

\appendix

\bibliography{apssamp}% Produces the bibliography via BibTeX.

\newpage

%\renewcommand\thefigure{A\arabic{figure}}    
%\setcounter{figure}{0} 

%\section{Experimental details}

%\begin{figure}[!ht]
%\centering
%\includegraphics[width=\linewidth]{fig/figA1_exp_crack.pdf}
%\caption{%
%\textbf{Comparison of fracture surfaces in non-optimized and optimized structures.}
%Post-fracture snapshots extracted from video frames show two representative non-optimized samples (left column) and four optimized ones (right). The magenta and blue contours highlight the segmented crack surfaces on the left and right sides of each specimen, respectively. While non-optimized structures exhibit relatively straight and localized cracks, optimized designs show more irregular and diffuse fracture paths, indicating improved stress redistribution and delayed failure due to topological optimization.
%}
%\label{fig:fracture_comparison}
%\end{figure}

\end{document}